# From WPT to Encrypted Telemetry: A Battery-Free Backscattering-based Polarimetric Wireless Sensor

Taki E. Djidjekh
LAAS-CNRS, Université de Toulouse, *CNRS*
Toulouse, France
taki.djidjekh@laas.fr

Loïc Thomas
LAAS-CNRS, Université de Toulouse, CNRS
Toulouse, France
loic.thomas@laas.fr

Quentin Bernyer
LAAS-CNRS, Université de Toulouse, CNRS
Toulouse, France
quentin.bernyer@laas.fr

Gaël Loubet
LAAS-CNRS, Université de Toulouse, CNRS, INSA
Toulouse, France
gael.loubet@laas.fr

Daniela Dragomirescu
LAAS-CNRS, Université de Toulouse, CNRS, INSA
Toulouse, France
daniela.dragomirescu@laas.fr

Alexandru Takacs
LAAS-CNRS, Université de Toulouse, CNRS, UPS
Toulouse, France
alexandru.takacs@laas.fr

***Abstract*—This work introduces an indoor Battery-Free Wireless Sensing Node powered through radiative Wireless Power Transfer (WPT). The proposed platform targets secure, energy-efficient active sensing and overcomes key limitations of many prior battery-free approaches, which commonly provide neither on-node computation nor cryptographic protection. The node combines temperature, humidity, pressure and Volatile Organic Compound (VOC) measurements with a low-power microcontroller that executes sensor calibration, derives a VOC index, formats the payload, and applies AES-128 encryption before wireless transmission. Energy harvesting and communication are enabled by a 1-bit controlled Backscatter Rectenna (BR), which both scavenges incident RF power and produces an orthogonally polarized backscattered signal for robust polarimetric operation. Experimental results validate reliable multi-sensor readout and encrypted data transfer, while maintaining a very low energy budget for the complete sense–compute–encrypt–transmit cycle.**

## I. Introduction

In the broader framework of Simultaneous Wireless Information and Power Transfer (SWIPT), battery-free wireless sensors and Wireless Sensor Networks (WSN) have become key enablers for long-term monitoring in structural health assessment, industrial instrumentation, and biomedical sensing scenarios [1]-[3]. Their appeal lies in removing maintenance-intensive process induced by the energy storage elements, while enabling distributed measurements in locations where wiring and periodic battery replacement are impractical.

Existing battery-free solutions can be broadly grouped into two families. A first class relies on fully passive transduction combined with RFID-like interrogation, radar-based sensing principles, or backscatter modulation [4], [5]. These architectures are attractive for their simplicity and the absence of active electronics at the node. Nevertheless, they typically remain confined to single-variable, low-complexity physical observables (e.g., temperature, humidity, strain), and the absence of meaningful embedded computation often forces all signal interpretation to be performed at the reader. From a cybersecurity standpoint, such passive devices generally provide little to no protection against practical threats—including impersonation/spoofing, eavesdropping, and replay/relay attacks—because neither robust on-node authentication nor cryptographic confidentiality is natively supported.

A second family consists of active Battery-Free Wireless Sensor (BFWS) nodes supplied either by ambient energy harvesting or by Wireless Power Transfer (WPT) from a dedicated RF source. While ambient sources are intrinsically variable in space and time, WPT enables a dedicated and therefore controllable RF energy supply, allowing more deterministic operation and duty-cycled sensing strategies. Active BFWS platforms typically integrate active sensors, measurement front-ends, and wireless communication circuitry, and can extend the operational range using standardized wireless technologies (e.g., RFID, BLE, LoRaWAN) or backscatter-based approaches. However, a common limitation in the current state of the art is that many active BFWS demonstrations still focus on a single sensing modality, with limited edge processing and minimal or absent encryption, leaving confidentiality and authenticity largely unaddressed [6]-[13].

This paper presents a low-power, microcontroller-based active BFWS powered by radiative WPT, designed for multi-parameter indoor environmental monitoring. The proposed node acquires temperature, humidity, pressure, and air-quality-related metrics, and executes local digital processing to condition the measurements and to compute an air-quality/VOC (Volatile Organic Compound) indicator directly at the sensor. Beyond sensing performance, the platform incorporates energy-aware identification and authentication at the physical layer to mitigate spoofing, relay/replay of frames or waveforms, eavesdropping, and man-in-the-middle attempts. Identification data, and the measured/processed environmental payload are then protected using AES-128 and transmitted through a polarization-domain backscatter mechanism. Specifically, an original 1-bit controlled rectenna simultaneously harvests RF power and maps the encrypted information onto a cross-polarized backscattered waveform, enabling robust polarimetric readout while maintaining a stringent energy budget.

The remainder of this paper is organized as follows. Section II describes the proposed BFWS architecture and its hardware/software implementation. Section III presents the experimental validation results. Section IV discusses the implications of the proposed approach. Finally, Section V concludes the paper.

## II. ARCHITECTURE AND IMPLEMENTATION

### A. System Hardware Architecture

Fig. 1 summarizes the proposed Battery-Free Wireless Sensor. The node is designed as an active, ultra-low-power sensing platform that operates only when sufficient RF energy is delivered by a dedicated WPT transmitter. It combines (i) a multi-parameter sensing front-end for indoor environmental monitoring, (ii) an MCU-based processing and security engine, and (iii) a dual-function RF interface that simultaneously enables energy harvesting and polarimetric backscatter communication through a 1-bit Backscattering Rectenna (BR). A dedicated Power-Management unit (PMU) conditions and buffers the harvested power to supply the digital and sensing subsystems.

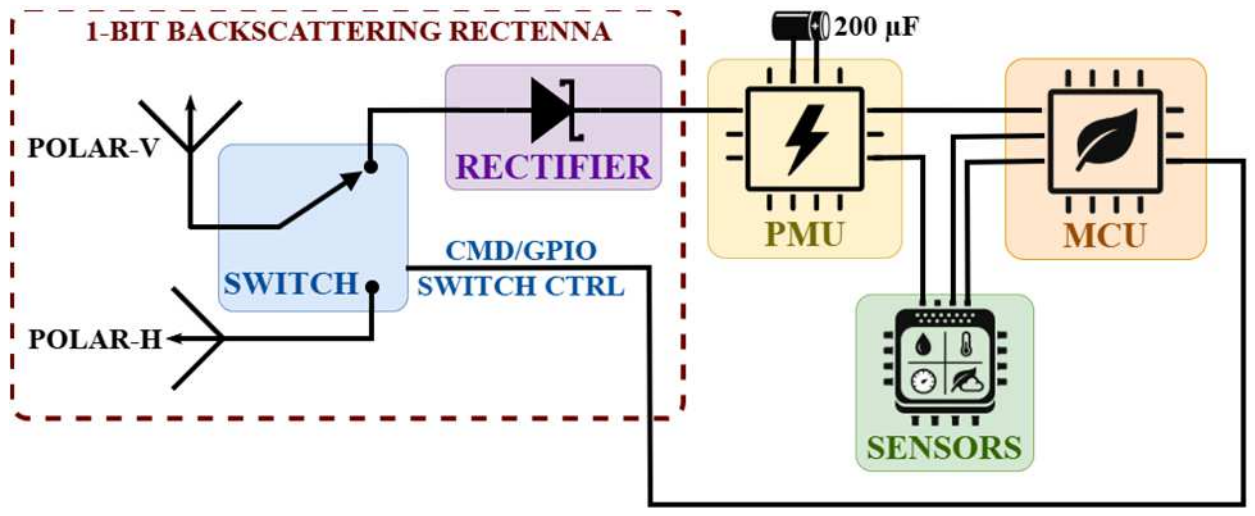

Fig. 1. Architecture of the proposed Battery-Free Wireless Sensor (BFWS).

#### 1) 1-bit Backscattering Rectenna

The RF front-end is implemented with two orthogonally polarized antennas (vertical and horizontal polarizations) interfaced to an RF rectification path through an RF fail-safe SPDT switch (GRF6011, Guerrilla RF) and an RF rectifier. The fail-safe behavior is exploited as a functional feature: in the absence of bias, the switch remains in its default low-loss state (insertion loss $\leq$ 0.4 dB), which prioritizes energy harvesting via the reception path (to the rectifier).

In this mode (Energy Harvesting), the incident vertical polarized WPT waveform is converted to DC and accumulated in the energy buffer through the PMU, allowing autonomous cold-start behavior.

Once the storage capacitor voltage is high enough to power the control electronics, the switch is biased at 3.3 V (typ. 1 mA supply current; 0.8 mA on the control pin). The BR then becomes digitally reconfigurable: a single-bit command (CMD) generated by an MCU GPIO toggles the RF state to implement backscatter modulation. In this active mode (Backscattering), the system transmits a modulated waveform on the cross-polarized (Horizontal polarized polar-H) antenna, providing polarization-domain separation between the WPT illumination (vertical polarization) and the backscattered telemetry waveform (horizontal polarization). This orthogonality reduces sensitivity to clutter and limits self-interference/cross-jamming between WPT illumination (downlink from RF source to BFWS) powering and uplink backscatter. The operation modes of BR are summarized in Table I.

TABLE I. OPERATING STATES OF 1-BIT BACKSCATTERING RECTENNA AS FUNCTION OF THE FAIL-SAFE RF SWITCH COMMAND.

| Harvesting Mode | Backscatter Mode |
|---|---|
| RF switch in fail-safe state (CMD/GPIO Low = 0 V) | RF switch MCU-controlled (CMD/GPIO modulated: ON = 3.3 V, OFF = 0 V) |
| POLAR-V antenna connected to RF rectifier | POLAR-V antenna connected to POLAR-H antenna |
| RF-to-DC conversion efficiency on 10 kΩ load: 16%–28% (depending on RF input power) | RF-switch insertion loss (POLAR-V input → POLAR-H output): 0.43 dB |

#### 2) The Power Management Unit (PMU)

The power-conditioning chain is realized with an energy-harvesting PMU (BQ25504, Texas Instruments) followed by a buck–boost DC/DC regulator (TPS63030, Texas Instruments). An 868 MHz RF rectifier and an energy buffer composed of two 100 µF (TL3B107K6R3C1700) storage capacitors (200 µF).

The BQ25504 supervises the charge/discharge window of the capacitors between 5.2 V (upper threshold) and 2.3 V (lower threshold), which corresponds to an available energy of approximately 2.2 mJ. The TPS63030 regulates the system rail to 3.3 V, supplying the sensing/processing electronics and, critically, the DC bias required for the RF switch during active backscatter operation.

#### 3) The Sensing Module and the Microcontroller Unit

The sensing front-end targets multi-modal indoor environmental observables. It integrates temperature and relative humidity sensing (SHT40, Sensirion), a metal-oxide VOC sensor (SGP40, Sensirion) providing a raw gas-related signal, and a barometric pressure sensor (LPS22DFTR, STMicroelectronics) to capture ambient pressure variations. All sensing elements are supervised over a shared low-power digital interface ($I^2C$), enabling tightly scheduled acquisitions under stringent energy constraints. In particular, the VOC measurements are interpreted using the manufacturer's gas index computation to deliver a normalized air-quality indicator.

Control, data handling, and security are implemented on an STM32G431KB MCU (Arm Cortex-M4, up to 170 MHz), and configured for low-power operation. The MCU orchestrates the full BFWS cycle: sensor triggering and readout, calibration/compensation and index computation, payload formatting, and AES-128 cryptographic processing. It also generates the CMD waveform that drives the 1-bit BR, thereby embedding protected information directly into the backscattered signal. The software strategy used to minimize the total energy per cycle is detailed next.

Fig. 2 presents a close-up view of the proposed BFWS node, highlighting the main functional building blocks. The prototype integrates the multi-parameter sensing module (temperature/humidity, VOC, and pressure), the low-power MCU, and the dedicated power-management stage with the storage capacitors. The RF front-end relies on a dual-polarized 1-bit Backscattering Rectenna. Together, these components implement the complete sense–compute–encrypt–transmit chain under radiative WPT. In this work, the proposed node architecture can be compacted onto a single PCB. However, for evaluation and proof-of-concept purposes, evaluation boards are used, and the RF rectifier and RF switch circuits are implemented on separate boards.

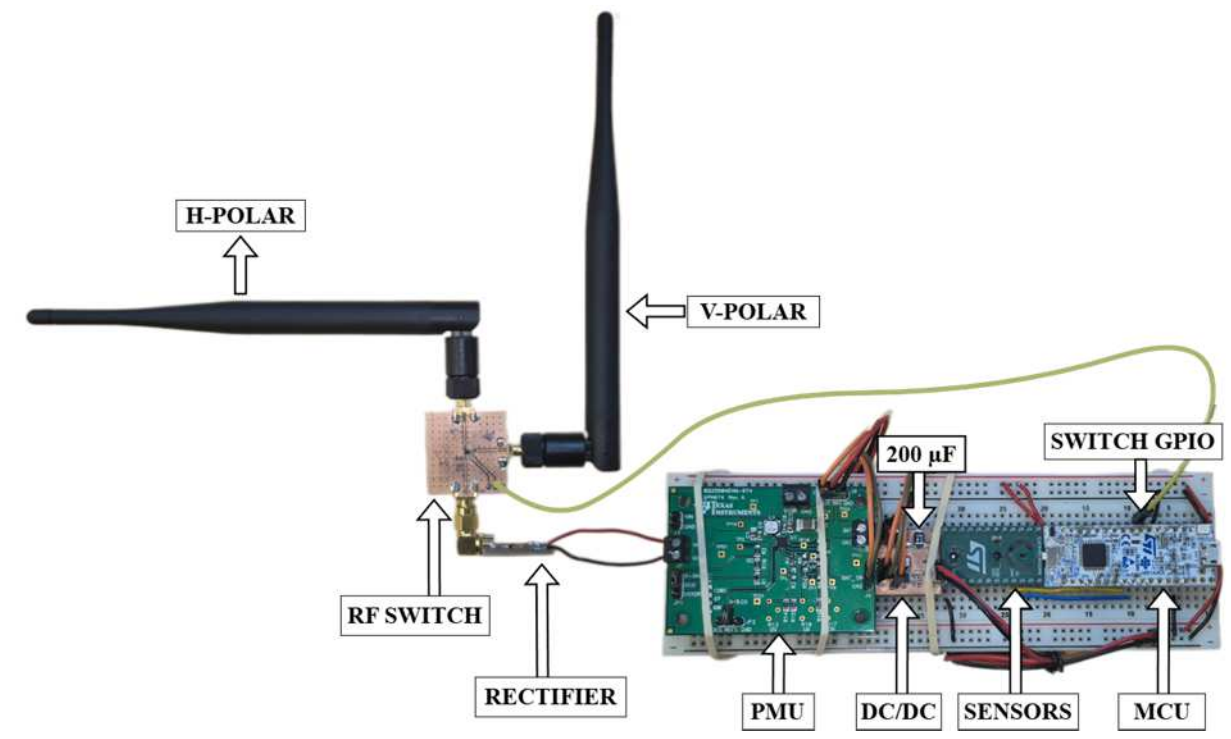

Fig. 2. Photograph of the proposed BFWS node prototype.

Fig. 3 shows a close-up view of the RF rectifier used for WPT energy harvesting. The 868-MHz input is delivered through the SMA connector and transformed by an impedance matching network (series inductor 33nH, reference LQW15AN33NG00H and shunt capacitor 4 pF, reference GRM1551X1H4R0CA01D both from Murata) to maximize RF-to-DC efficiency. Rectification is performed by the SMS7621-005LF Schottky diode, while a smoothing capacitor filters the output and stabilizes the DC level. The resulting DC voltage is routed to the PMU via the output leads.

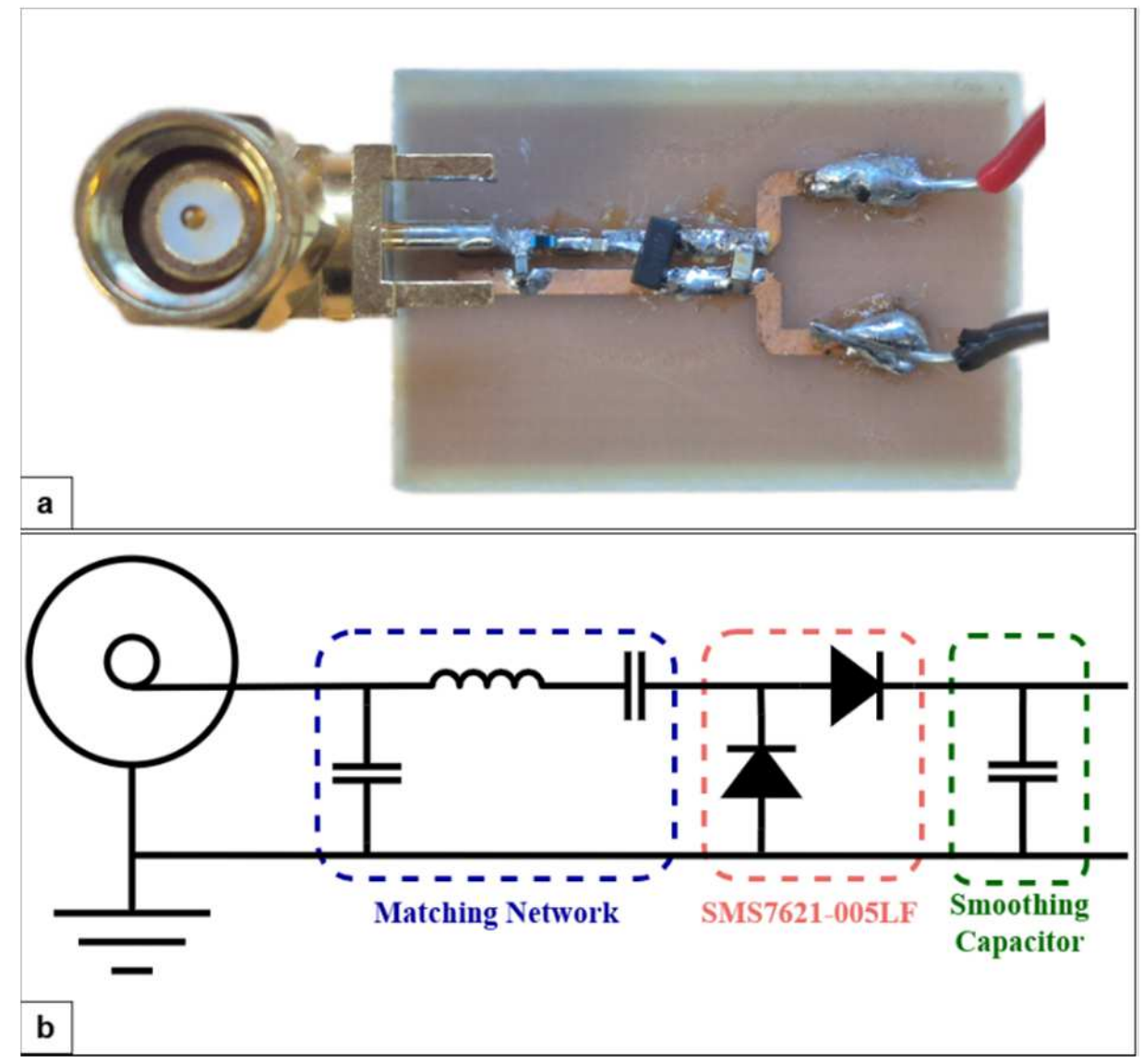

Fig. 3. (a) Photograph of the 868 MHz RF rectifier; (b) schematic of the RF rectifier circuit.

Fig. 4 reports the measured RF-to-DC conversion efficiency of the proposed rectifier versus input power (Pin) for $R_{load}$=10kΩ, across 864–880 MHz (around 868 MHz).

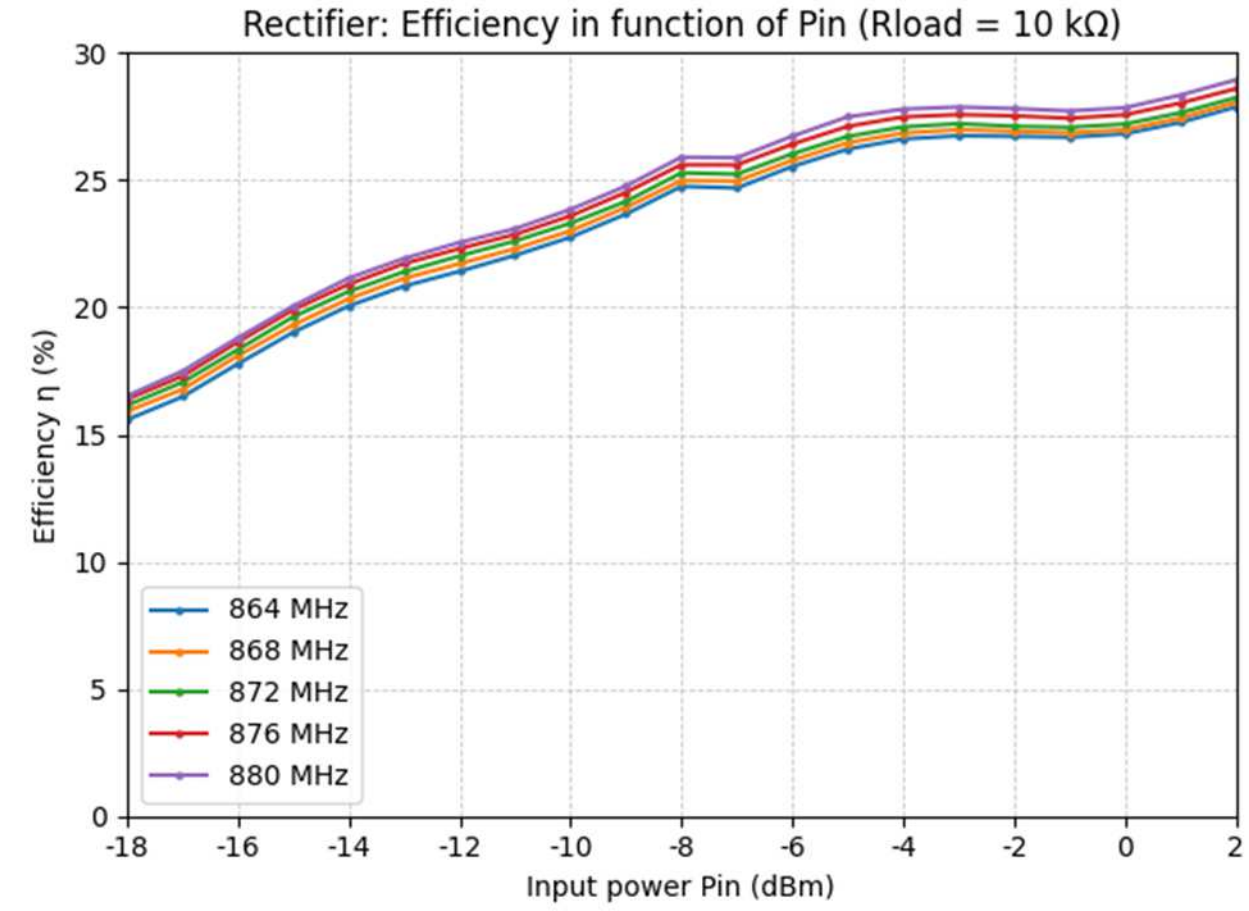

Fig. 4. Measured RF-to-DC power conversion efficiency of the rectifier versus RF input power with 10kΩ load.

The RF rectifier was optimized and experimentally tuned to efficiently operating by having the PMU as load that dynamically changes its input impedance as imposed by its own Maximum Power Point Tracking (MPPT). Nevertheless, the results depicted in Figure 4 confirm robust operation around 868 MHz and quantify the available conversion efficiency under low-input-power conditions.

## *B. Energy-Efficient System Software Implementation*

To reduce the global energy consumption, the MCU uses Dynamic Voltage and Frequency Scaling (DVFS) and operate in the (range 2) low power mode, with a 16 MHz frequency with the Phase Locked Loop (PLL) disabled. All sensors are accessed via I²C, and low-power sleep is managed through Real-Time Clock (RTC) support.

For air-quality estimation, the SGP40 provides a raw measurement that is processed by Sensirion's VOC Index algorithm [14]. This algorithm performs compensation, filtering, and baseline tracking to output an indoor air-quality metric between 0 and 500, where values around 100 correspond to stabilized good air quality and values approaching 500 indicate very poor air quality.

A key constraint is the VOC sensor's heater warm-up time (~170 ms). The proposed implementation therefore overlaps this warm-up with other tasks to avoid idle active time. After initializing the low-power configuration, configuring the BR control GPIO, and enabling I²C/RTC peripherals, the firmware first issues a dummy VOC command to start heater activation. During the warm-up interval, the MCU executes operations that do not require the final VOC sample: it retrieves and updates persistent state from flash (counter and algorithm parameters), performs flash loading/cleaning (22.16 ms), requests temperature/humidity acquisition from the SHT40 (13.2 ms), initializes the AES engine (0.1 ms), and measure pressure (2.9 ms) for a total active time of 38.36 ms. The remaining ~132 ms are spent in STOP1 with an RTC wake-up, substantially reducing energy consumption while the heater reaches operating conditions. Once warm-up is complete, the actual SGP40 measurement is performed; the heater current is approximately 2.6 mA. The described scheduling minimizes wasted energy by aligning compute and I/O tasks with the mandatory warm-up delay.

After reading the raw SGP40 output, the heater is disabled, and the VOC Index computation is performed under the low-power MCU configuration. The resulting data, including device identification, freshness protection, and environmental measurements—temperature, humidity, pressure, and air-quality metric—are then encapsulated into a compact 20-byte backscatter frame, whose structure is summarized in Table II.

TABLE II. UPLINK BACKSCATTER FRAME STRUCTURE

| Field | Size (bytes) | AES-128 | Description |
|---|---|---|---|
| Preamble | 2 | No | Synchronization pattern for frame detection |
| UID | 2 | Yes | Unique device identifier |
| Counter | 4 | Yes | Frame/iteration counter (anti-replay) |
| Data | 10 | Yes | Measured sensors payload |
| CRC | 2 | No | Bit-error detection |
| Total | 20 | - | Complete frame length |

The uplink frame consists of a 2-byte preamble, 2-byte UID, 4-byte counter, 10-byte sensor-data payload, and 2-byte CRC, and is transmitted using Manchester encoding. AES-128 encryption in ECB mode is applied to the UID, counter, and sensor-data fields, while the preamble and CRC remain unencrypted for synchronization and bit-error-detection purposes. A symmetric pre-shared key is stored in non-volatile memory, and the monotonic counter provides freshness protection by allowing the receiver to detect and reject replayed frames. Since the counter changes at each transmission, the encrypted plaintext block is updated for every frame, reducing ciphertext repetition in normal operation. Spoofing resistance is supported by the combination of encrypted device identification, counter validation, and key secrecy. Finally, the VOC algorithm state and iteration counter are committed to flash memory to preserve continuity across intermittent energy cycles.

With this duty-cycled execution and power-management approach, a complete sensing, processing, encryption, transmission, cycle requires 1.83 mJ at 3.3 V, as measured with the Nucleo-LPM01A. The PMU gives enough energy for powering a whole cycle. Fig. 5 reports the measured current trace for a full cycle with task-level annotations, highlighting that the dominant contribution stems from the VOC warm-up overhead.

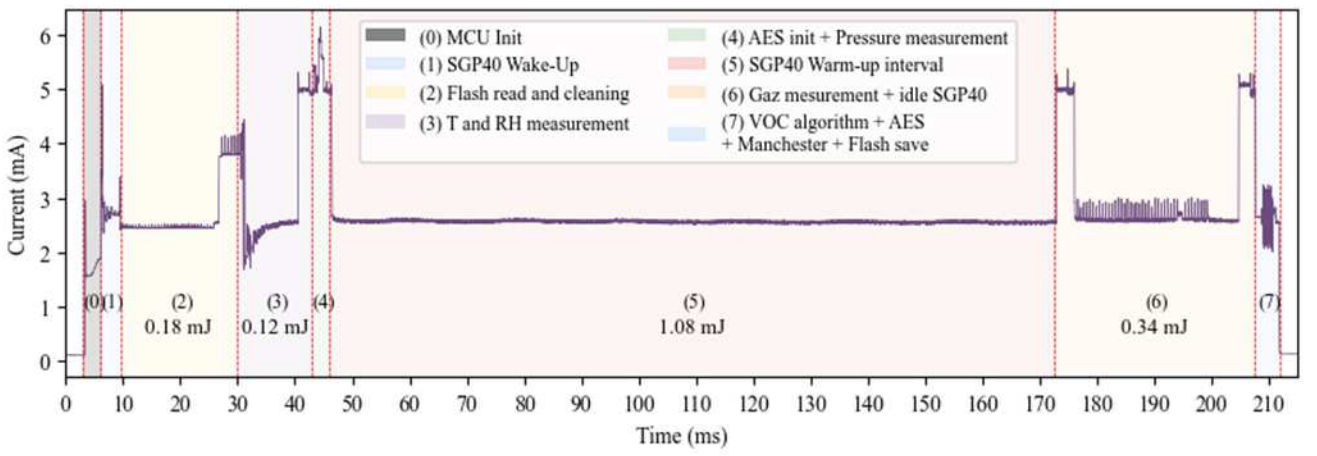

Fig. 5. Measured current consumption at 3.3V over a complete cycle with annotated tasks.

## III. EXPERIMENTAL VALIDATION

### A. Measurement Setup

The proposed BFWS was assessed in a representative indoor environment using a purpose-built measurement bench. The WPT illumination was generated by an Anritsu MG3694B RF signal generator at 868 MHz and adjusted to 30 dBm EIRP, radiated through a vertically polarized (V-pol) patch antenna. The uplink telemetry was observed on a separate receive chain composed of a Tektronix RSA306B USB spectrum analyzer connected to a horizontally polarized (H-pol) monopole, thereby selectively capturing the cross-polarized backscattered waveform. The BFWS prototype—integrating the dual-polarized rectenna and the sensing/processing electronics—was placed at a separation distance between 1.50 meters to 7 meters from the WPT source (Fig. 6).

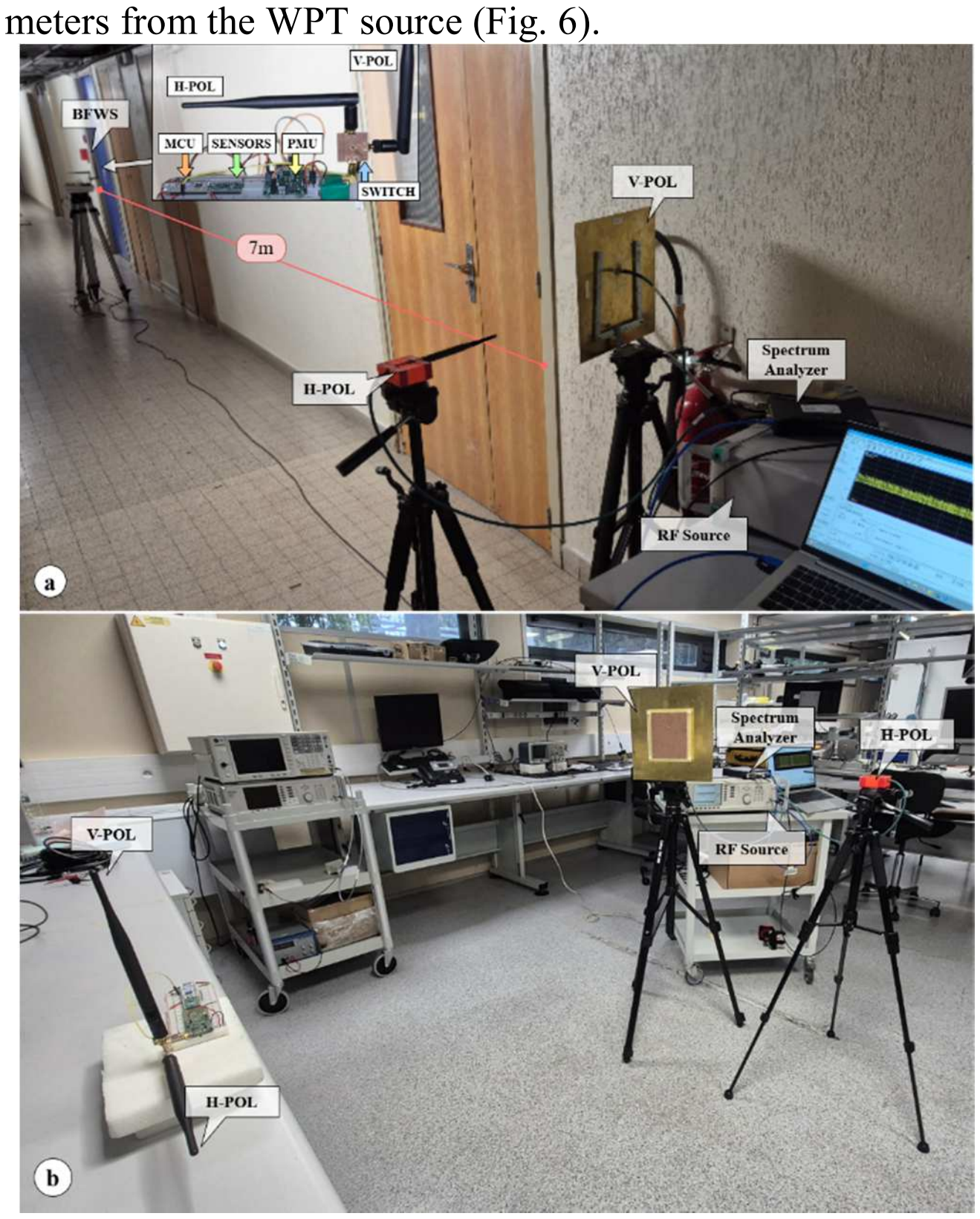

Fig. 6. Indoor experimental setup: 868 MHz WPT RF source (V-pol) and spectrum-analyzer receiver (H-pol), with the BFWS prototype positioned at (a) 5–7 m (zoomed view of the BFWS node) and (b) 1.5–3 m.

### B. Functional Demonstration and Data Recovery

Under continuous RF powering, the node operated in a stable duty-cycled regime and periodically executed a complete sense–process–encrypt–transmit sequence. The BFWS successfully reported temperature, relative humidity, barometric pressure, and the VOC-derived air-quality indicator at a ~52 seconds interval (+33 dBm EIRP, 2 meters distance), demonstrating sustained multi-sensor operation under radiative WPT. The received backscatter waveform was recorded by the spectrum analyzer and decoded to reconstruct the transmitted packet structure.

Fig. 7 illustrates a representative captured frame, with byte-level delineation of the payload. To verify end-to-end integrity of the secure telemetry chain, the encrypted payload was subsequently processed offline using a Python-based post-processing script implementing the same cryptographic material as the node. The resulting recovered after decryption match the expected formatting and contain the reconstructed environmental quantities (temperature, humidity, pressure, and VOC index), confirming correct operation of the on-node encapsulation, AES-128 protection, and polarization-based backscatter embedding.

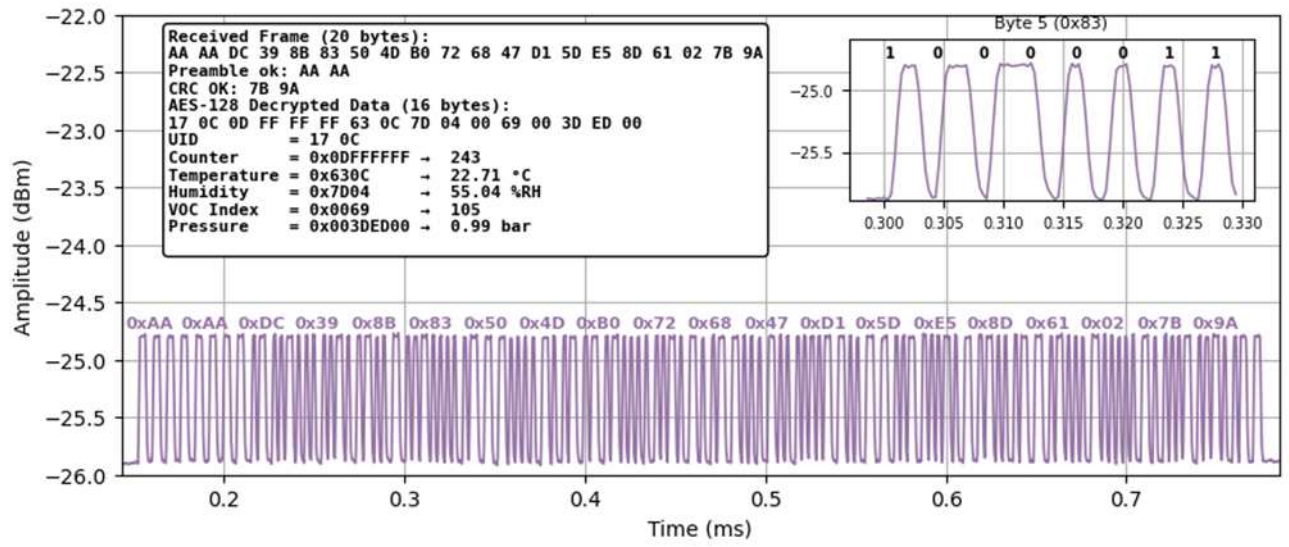

Fig. 7. Received cross-polarized backscatter frame captured by the spectrum analyzer, with annotated packet parsing, zoomed view of the fifth byte, and recovered sensor values after AES-128 decryption.

In addition to functional sensing validation, the WPT powering capability was characterized through charging-time measurements in two representative indoor scenarios. Table III reports the charging time as a function of distance at the maximum EIRP setting (868 MHz, +33 dBm). The energy buffer reached the regulation threshold in 3 minutes 11 seconds at 5 meters, and in 5 minutes 35 seconds at 7 meters, indicating that the proposed architecture remains operable at meter-scale separations and suggesting feasible operation at even longer distances under favorable propagation conditions. Table IV complements this analysis by presenting the charging time versus RF source power level in an office environment, highlighting the expected sensitivity of the replenishment time to transmit-power variations and providing practical guidance for selecting EIRP levels to meet a target measurement duty cycle.

TABLE III. CHARGING TIME VS DISTANCE IN A CORRIDOR (33DBM EIRP)

| **Distance** | 5m | 7m |
|---|---|---|
| **Charging time**[b] | 191s [a] | 335s [a] |

[a] Measurement results in the corridor differ from those in the measurement room due to the difference in wave reflection.

[b] The choice of the corridor, due to the distance constraint, makes the environment more favorable to the WPT.

TABLE IV. CHARGING TIME (S) VS EIRP LEVELS (IN AN OFFICE)

| **EIRP →** | **+33 dBm** | **+32 dBm** | **+31 dBm** | **+30 dBm** | **+29 dBm** |
|---|---|---|---|---|---|
| 1.50 m | 15 s | 20 s | 25 s | 32 s | 45 s |
| 2 m | 52 s | 69 s | 95 s | 142 s | 217 s |
| 2.50 m | 82 s | 132 s | 263 s | X [a] | X [a] |
| 3 m | 234 s | X [a] | X [a] | X [a] | X [a] |

[a] Charging times exceeding 4 minutes are considered not optimum for the VOC measurement algorithm.

## IV. DISCUSSION

The experimental results obtained in a real indoor environment confirm the correct end-to-end operation of the proposed WPT-energized Battery-Free Wireless Sensing (BFWS) node, including sensing, local processing, payload construction, AES-128 protection, and polarimetric backscatter telemetry. The platform demonstrates that multi-parameter environmental sensing—temperature, relative humidity, barometric pressure, and VOC-related air-quality monitoring—can be combined with embedded security and backscatter-based communication within a stringent energy budget.

A complete sensing, calibration, and secure transmission cycle was measured at 1.83 mJ on the 3.3 V rail, demonstrating practical feasibility for duty-cycled operation under radiative WPT. The charging-time characterization also provides system-level insight into the available powering margin. In corridor measurements at a maximum EIRP of +33 dBm, the measured charging time was 3 min 11 s at 5 m and 5 min 35 s at 7 m, indicating that the architecture can support operation over several meters. Complementary office measurements further quantified the dependence of recharge time on RF source power, providing a practical basis for tuning the transmit power level according to a target sampling period.

The communication data rate used in this work was set by the GRF6011 switch modulation frequency. With a switching frequency of 400 kHz and Manchester encoding, the effective frame data rate is 200 kbps, since each information bit is represented by two modulation states. For the 20-byte frame, corresponding to 160 information bits, the resulting frame duration was 0.8 ms.

This data rate is sufficient for the compact sensing payload considered in this work. However, the RF switch is capable of switching at up to 6 MHz, which could significantly increase the achievable backscatter data rate in future implementations, provided that the MCU timing, receiver bandwidth, signal-to-noise ratio, and demodulation chain are adapted accordingly.

Overall, the results show that the proposed architecture advances battery-free sensing beyond common single-sensor and reader-centric paradigms by combining multi-sensor acquisition, embedded processing, lightweight security, and polarization-separated backscatter communication in a WPT-powered node.

## V. CONCLUSION

This paper presented a WPT-energized Battery-Free Wireless Sensing node that combines multi-parameter environmental monitoring, embedded security, and polarimetric backscatter telemetry. The proposed platform integrates active temperature, relative humidity, barometric pressure, and VOC-related air-quality sensing, supervised by a low-power MCU that performs local calibration, feature extraction, payload construction, and AES-128 ECB protection before transmission.

Uplink communication is achieved through an original 1-bit polarization-shift rectenna controlled by the MCU and operating in two modes: energy harvesting and backscattering. In backscatter mode, the protected data are embedded into a cross-polarized backscattered waveform, providing polarization-domain isolation between the powering waveform and the telemetry channel. This separation improves robustness against indoor clutter and cross-interference.

The experimental validation confirmed the practical feasibility of the proposed BFWS architecture under radiative WPT, with reliable sensing, secure processing, and backscatter-based communication demonstrated in indoor conditions. Future work will focus on increased integration and miniaturization, including implementation on a compact single-PCB architecture, adoption of a single dual circularly polarized antenna to enhance robustness against clutter and cross-jamming, and further optimization of the energy efficiency and communication performance.

## Acknowledgement

This research was funded, in whole or in part, by the French National Research Agency (ANR) under the project SWAVE "ANR-25-CE39-5853", and the authors gratefully acknowledge its support.